# ELEMENTS OF LEGISLATION FOR ARTIFICIAL INTELLIGENCE SYSTEMS

Anna Romanova

Moscow Institute of Physics and Technology (National Research University), Moscow, Russia

*ABSTRACT*

*The significant part of the operational context for autonomous company management systems is the regulatory and legal environment in which corporations operate. In order to create a dedicated operational context for autonomous artificial intelligence systems, the wording of local regulatory documents can be simultaneously presented in two versions: for use by people and for use by autonomous systems. In this case, the artificial intelligence system will get a well-defined operational context that allows such a system to perform functions within the required standards. Local regulations that provide basis for the joint work of individuals and autonomous artificial intelligence systems can form the grounds for the relevant legislation governing the development and implementation of autonomous systems[1].*

## 1. INTRODUCTION

Artificial intelligence systems that manage corporations must work effectively not only with objects of the material world, but also in the legal environment [1]. "At least since Leibniz, the dream of excluding man from the spiral of legal reasoning has captured the imagination of philosophers, lawyers and (more recently) computer scientists" [2]. Leibniz's idea is presented "in his Dissertatio de Arte Combinatoria as Universal Mathematics, a theoretical, formal system of propositions and rules that would allow all disputes to be resolved with mathematical precision" [2]. Historically, laws were created and enforced by people. With the development of artificial intelligence technologies, laws will be executed by machines. The level of wording precision acceptable by modern humans is much lower than the level of wording precision required for an artificial system. There is a fundamental difference "between human decisions, as social constructs, and algorithmic decisions, as technical constructs" [3]. Stephen Wolfram speaking at SXSW 2013 said: "computing will become central to almost every field" [4]. Wolfram believes that: "we are now almost ready ... for a computational law. Where, for example, contracts become computational. They explicitly become algorithms that decide what is possible and what is not" [4]. Wolfram suggests that it will be necessary to adopt a separate constitution for artificial intelligence systems (hereinafter referred to as AI) [5]. However, the question is: "What should be in such a constitution?" [5] remains open for now. The paper presents proposed foundations for AI constitution: in the section two the operational context and operational design domain are discussed. In the section three the establishment of a dedicated operational context for autonomous management systems is presented. The section three also discusses the proposed practical steps to develop AI constitution.

## 2. OPERATIONAL CONTEXT AND OPERATIONAL DESIGN DOMAIN

The most effective approaches to the development and application of civil and commercial autonomous systems have currently been developed in the field of autonomous vehicles. In course of developing autonomous vehicles, the concept of "operational design domain (ODD)" [6] is used, which "is an abstraction of the operational context, and its definition is an integral part of the system development process" [6]. The Society of Automotive Engineers (SAE) International Standard J3016 "Taxonomy and Definitions of Terms Associated with Driving Automation Systems for On-Road Vehicles" defines the operational design domain as "the combined operating conditions under which a given driving automation system (or function thereof) is specifically designed to operate" [7]. "It is necessary to know the operational context to provide performance and safety assurance" [6]. And "the required level of security is only guaranteed in a well-defined and tested operational design domain" [6]. British Standards Institution (BSI) states that "a key aspect of the safe use of an automated vehicle is identifying its capabilities and limitations and communicating this clearly to the end user, resulting in a state of "informed safety" [8]. The British Standards Institution believes that "the first step in establishing the capabilities of automated vehicles is to define the Operational Design Domain (ODD)" [8].

For autonomous corporate governance systems a significant part of the operational context are laws and other regulations, as well as the interpretation and application of laws and regulations by other AI systems and humans. Stephen Wolfram is not the only one who has come to the conclusion that significant changes of the legal system are requiredto enable autonomous AI systems operate effectively. The European Commission's report on the "Ethics of Connected and Automated Vehicles" states that autonomous systems will not be able to follow the rules created for humans precisely. For the successful implementation of self-driving cars, several options must be considered: "(a) traffic rules must be changed; (b) autonomous cars should be allowed to disobey traffic rules; or (c) autonomous cars must relinquish control so that a person can decide not to obey traffic rules" [9].

Modern legal practice already has examples when an additional or special operational context is created for systems with a significant difference in the worldview — these are bilingual contracts. For example, "Chinese law requires that a joint venture agreement be approved by Chinese government authorities" [10]. Therefore, it is natural that "the joint venture agreement should be written in Chinese" [10]. In those cases where a dual operational context is needed, "the joint venture agreement is a bilingual agreement: there is one agreement, but with two different texts, one in English and one in Chinese" [10].

Many researchers still formulate possible criteria for a decision based on the usual capabilities of human abilities: "race, religion, sex, disability, age, nationality, sexual orientation, gender identity or gender expression" [11], while describing situations where an autonomous system will have to make decisions that are not obvious to humans. Researchers from the Massachusetts Institute of Technology (MIT) in the famous Moral Machine experiment also used factors only understandable for humans: gender, age, etc. [12]. For AI systems based on mathematical algorithms such criteria have no meaning.

In the modern world there are already AI systems that autonomously solve specific problems without directly making management decisions. These are systems that make very fast financial decisions—algorithmic and high-frequency trading systems [13]. Special regulatory techniques have been developed for such systems: "information disclosure, internal testing and monitoring systems" [13]. For such systems "structural features of the trading process" are provided [13], i.e. a special operating context has been created for such systems.

## 3. Establishing an Operational Context for Autonomous Management Systems

By analogy with the J3016 standard "Taxonomy and definitions of terms associated with driving automation systems for road vehicles" we propose to formulate for autonomous corporate management systems the respective concept of an operational design domain: these are the combined operating conditions under which a given control automation system (or its function) is specifically designed to function. Most of the operational context for autonomous vehicles are physical objects, but for autonomous corporate management systems "the regulatory and legal environment within which corporations operate is critical to overall economic performance" [14]. A significant part of the operational context for autonomous corporate governance systems consists of a variety of regulations [1]. The G20/OECD Principles of Corporate Governance state that "the objectives of corporate governance are also formulated in voluntary codes and standards that do not have the status of law or regulation" [14]. In order to create a specific operating context for autonomous AI systems, the internal policies can be presented simultaneously in two versions: for use by humans and for use by autonomous systems. In this case, the AI system is given a clearly defined operational context that allows the system to perform functions within the required operational capabilities [1].

The basic principles of corporate governance, as well as the basic functions of the board of directors, are set out in the G20/OECD Principles of Corporate Governance [14]. In order to clarify and implement the G20/OECD Principles many countries and companies are developing and applying their own, more detailed, corporate governance codes. The G20/OECD principles and corporate governance codes form the grounds that provides the basis for creating the operational context for autonomous corporate governance systems. Let's look at some examples of policy wording for autonomous AI systems in mixed boards of directors (boards consisting of individuals and autonomous AI systems).

A key principle of corporate governance is "fair treatment of all shareholders" [14]. The concept of fair treatment for autonomous systems in modern practice is formalised using the principles of informed consent [9], non-discrimination [11], and fair statistical risk distribution [9].

### 3.1. Informed Consent

Modern AI researchers conclude that "engineers do not have the moral right to make ethical decisions on behalf of users in difficult cases when the stakes are high" [15]. The report on "The Ethics of Connected and Automated Vehicles" states that for the use of autonomous systems, it is necessary to develop "more nuanced and alternative approaches to user agreements" [9]in order to obtain informed consent, rather than simply a "take it or leave it" approach [9]. Informed consent involves informing the user about how the AI system will behave under normal conditions and in critical situations. Road accidents involving Tesla autopilots show that "it is unclear whether Tesla beta testers were fully informed of the risk. Did they know that death was possible?" [16]. Corporate policies, regulations, and codes that describe the principles and rules of operation of an autonomous AI system will allow shareholders and other stakeholders to express informed consent for the use of such a system in corporate governance. In our view, corporate governance rules must be executed simultaneously by both individual directors and autonomous systems and such rules can be compiled in two editions — for individuals and for autonomous AI systems:

- Policies, regulations, and codes for individuals should govern issues based on the individual's mindset;
- Policies, regulations, and codes for autonomous AI systems should address corporate governance issues based on the metrics available to AI systems.

### 3.2. Non-Discrimination

The report on "The Ethics of Connected and Automated Vehicles" states that "discriminatory service delivery" [9] by autonomous systems must be avoided. An AI system can and should be tested for bias, direct and indirect discrimination [17]. The corporate policy written for a mixed board of directors should specify which tests the autonomous system must pass, their frequency, and a list of signs of direct and indirect discrimination.

The G20/OECD Principles state that "the board should apply high ethical standards" [14]. The report on "The Ethics of Connected and Automated Vehicles" notes that in critical situations, "it may be impossible to regulate the exact behavior" [9] of autonomous systems. Therefore, the EU expert group proposes that the behavior of autonomous systems should be considered ethical if "it emerges organically from a continuous statistical distribution of risk by the CAV in the pursuit of improved road safety and equality between categories of road users " [9]. For a fair distribution of risk, modern researchers are trying to use abstract values: "a proportional relationship between the velocities of the traffic participants and the severity of harm can be established independently of any ethical evaluation" [18]. This approach is not always possible, especially in the case of limited resource allocation. In practice, several algorithms have been developed for the allocation of scarce medicines "treating people equally, favouring the worst-off , maximising total benefits, and promoting and rewarding social usefulness" [19]. Therefore, policies, codes, and regulations drawn up for a mixed board of directors should disclose to shareholders and third parties how the requirements for autonomous systems are formed, namely, the criteria and indicators for the formation of algorithms for fair distribution of risks.

### 3.3. Monitoring Results of Management Activities

The G20/OECD principles determine that "board is chiefly responsible for monitoring managerial performance" [14]. Situations may arise when an autonomous AI system evaluates the performance of a human manager. The Portuguese Corporate Governance Code states that "the non-executive directors shall exercise, in an effective and judicious manner, a function of general supervision and of challenging the executive management" [20]. Moreover, similar to the concepts of continuous reporting and continuous auditing, an autonomous AI system is capable for "continuous monitoring" of the results of management activities. The concept of "continuous auditing" [21] was proposed in 1991 at AT&T Bell Labs for auditing large digital databases [21]. The system "encompassed the monitoring and real time assurance on a large billing system focusing of the data being measured and identifying through analytics methods faults in the data that lead both to control and process diagnostics"[22]. The authors of the concept indicated that its implementation will require "major changes in software, hardware, the control environment, management behavior, and auditor behavior" [21], i.e. changing the existing operational context or creating an additional one. The concept of continuous auditing is inextricably linked with the concept of "continuous reporting" [22]. When creating policies and procedures for autonomous AI systems, a company should determine whether it is gaining a significant competitive advantage by tracking transactions, events, and information in real time. Policies, codes, and regulations for autonomous AI systems must contain a specific list of activities, sources, and implications for monitoring the effectiveness of management activities.

## 3.4. Legal Compliance

The G20/OECD Principles state that the responsibilities of the board of directors include oversight of "the risk management system and mechanisms designed to ensure that the corporation obeys applicable laws" [14]. Modern regulatory risk monitoring systems can provide continuous risk monitoring in many areas simultaneously: "anti-bribery and corruption regulatory compliance, anti-money laundering compliance, financial services regulatory compliance, risk assessments of current and prospective business partners, agents and vendors, mergers and acquisitions and other investments in emerging and global markets, industry and jurisdiction-specific risk" [23]. To achieve such broad and detailed analysis, companies create systems that "consolidates data from a wide range of worldwide data sources" [23]. The list of data sources that an autonomous system can collect and analyze could be very diverse: leading data aggregators, screening in media and / or judicial reviews, information about corporate structure and operations, third party holdings and shareholders [23]. Corporate policies, codes, and regulations drawn up for autonomous AI systems in a mixed board of directors should contain a specific list of sources and schedule for information updates.

## 3.5. Considering Interests of Third Parties

In disclosing the responsibilities of boards of directors, the G20/OECD Principles specify that "they are expected to consider and treat fairly the interests of stakeholders, including employees, creditors, customers, suppliers and affected communities" [14]. The European Commission's report on the "Ethics of Connected and Automated Vehicles" suggests that autonomous systems should "adapt their behaviour around vulnerable road users instead of expecting these users to adapt to the (new) dangers of the road" [9]. It also suggests that autonomous systems "should also be designed in a way that takes proactive measures for promoting inclusivity" [9]. Researchers from the Technical University of Munich propose to include special parameters for less protected users in the algorithm for fair risk distribution [18]. Corporate policies, codes, and regulations written for autonomous AI systems as part of a mixed board of directors, should contain indicators that have to be taken into account when considering the interests of third parties.

## 3.6. Awareness, Integrity, Prudence, and Care

The G20/OECD Principles state that "board members must act in a fully informed manner, with integrity, due care and diligence, and in the best interests of the company and shareholders" [14]. For AI systems, awareness is formalized on the list and scope of necessary sources and data, as well as the regularity of updating sources, data, algorithms and models. When there is a significant volume of transactions, it is impossible for a person to establish a duty to verify every transaction at every point in time. The AI system can verify transactions in real time or at a certain interval [24], [25]. The AI system can check all operations, or only certain ones [24], [25]. For an AI system, it is possible to prescribe the list of sources that it will use [24]. By analogy with the concepts of continuous reporting and continuous auditing, a company may consider implementing "continuous awareness" and "continuous monitoring". Corporate policies, codes, and regulations drawn up for autonomous AI systems as part of a mixed board of directors should answer the following questions: what is the list sources of information, what is the regularity of updating sources, data, and models, what is the catalog of necessary activities.

### 3.7. Appointment of a Director

G20/OECD Principles suggest that when appointing an individual as a director, consideration should be given to his or her "relevant knowledge, competence and experience " [14]. For example, the Saudi Arabian Corporate Governance Code states that information about candidates nominated for the position of director must disclose "experience, qualifications, skills and their previous and current employment and memberships" [26]. It is also required that the candidate "must have academic qualifications and appropriate professional and personal skills, as well as an appropriate level of training and practical experience" [26]. Currently, many large banks prohibit employees from using the ChatGPT system for business purposes due to its "inaccuracy and legal problems" [27]. Corporate policies, codes, and regulations drawn up for an autonomous system as part of a mixed board of directors should answer the following questions: by what parameters the AI system is selected, what tests or exams it must pass.

### 3.8. Performance Evaluation

The G20/OECD Principles state that "boards of directors should regularly assess their performance and determine whether they have the required combination of experience and competencies" [14]. The G20/OECD principles suggest that "with the help of training" [14] board members can maintain the required level of knowledge. Anautonomous system does not require courses and training, but regular updating of data and algorithms. Therefore, corporate policies, codes and regulations drawn up for autonomous AI systems as part of a mixed board of directors must answer the following questions: how often, to what extent, and based on what sources, algorithms and data should be updated.

### 3.9. Cooperation

It is necessary to provide methods for effective communication between autonomous AI systems and other stakeholders (directors , shareholders , managers , employees , etc.) in terms of mixed board of directors. Individuals usually work in accordance with the work schedule. Such periodicity is justified for distributing the effective workload for individuals, but does not make sense for determining the operating mode of an autonomous AI system that can work around the clock. The very concept of business meetings and effective communication is also changing. An autonomous AI system will use a digital interface to communicate: "chatbots (for example, conversational AI via audio or text), visual holograms, virtual or augmented reality" [28].

Best practices for specifying dedicated operational context for autonomous AI systems can be summarised and used in legislative activities . One would expect that individuals would also prefer to use the more precise policies developed for autonomous systems, but individuals would not be able to process the volume of data required.

## 4. CONCLUSIONS

Lawmakers, developers, and researchers in the field of artificial intelligence are faced with a choice: whether it is necessary to create special conditions for the functioning of autonomous artificial intelligence systems or whether they can function in the same operational context as ordinary people. In other words, can autonomous systems be considered like a car that can be used on a public highway, or is it more like a train, plane, or rocket, and requires a dedicated infrastructure to effectively use such systems.

Analysis of emergency situations in transport [29] clearly shows that despite the incomparable power and speed, air transport and railway transport are several times safer than cars. Huge and safe speed and power for planes and trains are achieved by dedicated infrastructure: railways, stations, airports and air corridors. The same principle of increasing secure efficiency through dedicated infrastructure is already partially applied in the field of algorithmic and high-frequency trading and can be applied to other autonomous AI systems.

A significant part of the corporate governance infrastructure is created in the form of internal company regulations. Internal regulations for joint work of individuals and autonomous AI systems can form the basis for the respective general legislation.

**AUTHOR**

**Anna Romanova**, Ph.D. in Economics. An expert in the field of digital transformation, Ph.D., MBA, LL.M, ALM, over 15 years of experience in the leading international companies. Postgraduate student at MIPT in the field of Artificial Intelligence and Machine Learning.

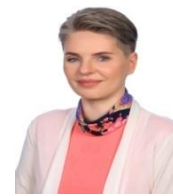